%% file: main.tex
\documentclass[a4paper, 14pt]{extarticle}
\usepackage{amssymb}
\usepackage{amsmath}
\usepackage[utf8]{inputenc}
\usepackage[OT1]{fontenc}
\usepackage{amsfonts}
\usepackage{ifthen}
\usepackage{tabularx}
\usepackage{textcomp}
\usepackage{physics}
\usepackage{listings}
\usepackage{color}
\usepackage{enumitem}
\usepackage{indentfirst}
\usepackage{csquotes}
\usepackage{fancyhdr}
\usepackage{setspace}
\usepackage{graphicx}
\usepackage{colortbl}
\usepackage{listings}
\usepackage{color}
\usepackage{booktabs}
\usepackage{geometry}
\usepackage{svg}
\usepackage{enumitem}
\usepackage{indentfirst}
\usepackage[colorlinks,citecolor=blue,linkcolor=blue,bookmarks=false,hypertexnames=true, urlcolor=blue]{hyperref} 
\usepackage{indentfirst}
\usepackage{mathtools}
\usepackage{booktabs}
\usepackage[flushleft]{threeparttable}
\usepackage{tablefootnote}
\usepackage{float}
\usepackage[justification=centering]{caption}

\definecolor{dkgreen}{rgb}{0,0.6,0}
\definecolor{gray}{rgb}{0.5,0.5,0.5}
\definecolor{mauve}{rgb}{0.58,0,0.82}

\usepackage{chngcntr}
\counterwithin{table}{section}
\counterwithin{figure}{section}

\usepackage{amsmath,amssymb,amsfonts}
\usepackage{algorithmic}
\usepackage{graphicx}
\usepackage{textcomp}
\usepackage{xcolor}
\def\BibTeX{{\rm B\kern-.05em{\sc i\kern-.025em b}\kern-.08em
    T\kern-.1667em\lower.7ex\hbox{E}\kern-.125emX}}
    
\usepackage[backend=biber, style=ieee]{biblatex}
\begin{document}
\input{title.tex}
\tableofcontents
\newpage
\section*{Abstract} 
\addcontentsline{toc}{section}{Abstract}
Many current applications have to perform data processing in a streaming fashion. Doing so at a large scale requires a parallel system that must be equipped to handle straggling workers and different kinds of failures. YT is the main driver behind distributed systems at Yandex, home to its distributed file system, lock service, key-value storage, and internal MapReduce platform. We implement a new component of this system designed for performing streaming MapReduce operations, utilizing different core YT solutions to achieve fault-tolerance and exactly-once semantics while maintaining efficiency and low write amplification factors.

\section*{Keywords}
streaming data processing, map-reduce, write amplification, fault-tolerance, exactly-once, distributed systems
\pagebreak

\section{Introduction}
In the modern world, many large-scale data processing problems require handling a continuous and quickly shifting stream of data. This is especially true at a company as prominent as Yandex, which is handling exabytes of data from web search and its other services.

Classic \textit{batch} approaches involve running many parallel computations on a static and usually large pool of data. Batch processing systems have been flourishing since the introduction of the MapReduce paradigm \cite{mr} by Google. However, real-time streaming conditions impose a certain amount of limitations for which the batch programming model is not a fit. 

The foremost consideration is that real-time processing on a stream of data demands much lower latencies than batch executions, which can typically take many hours. This implies that persisting data is now much more costly relative to the execution time but still needed due to potential failures. It is thus an overly important task to reduce \textit{write amplification}, which is the phenomenon associated with the same data being written to storage multiple times. Consequently, it is essential to keep as much data \textit{in memory} as possible, so all of the consumers of a stream have to move along at a similar speed. This makes \textit{straggling workers} even more of a challenge than before.

We implemented a streaming version of MapReduce optimized to handle the aforementioned problems. It is part of Yandex's YT system and heavily utilizes its other components.

Though there have been many developments in this field over the past decade, we think that taking advantage of Yandex's unique infrastructure will benefit internal users and external services greatly, while also bringing something new to the table of streaming data processing.

This thesis is structured as follows: a broad synopsis is given in the subsections below, chapter \ref{section:rw} discusses related work in this area, chapter \ref{section:yt} gives a short overview of the YT components used; chapter \ref{section:sa} describes the system architecture, chapter \ref{section:evaluation} evaluates the results, chapter \ref{section:fw} proposes designs for several future enhancements and chapter \ref{section:conclusion} concludes and summarizes the work.

\subsection{Relevance and significance}
\label{subsection:ras}
Big companies like Google, Yandex, Facebook, Amazon, Twitter and others deal with ginormous amounts of data, currently on the order of billions of gigabytes. Even smaller companies need to analyze and act on different kinds of logs and metrics from their services, often using cloud services which provide tools for efficient data processing, such as Amazon Web Services or the Google Cloud Platform.

In all of the cases above doing the processing on a single machine becomes economically unfit. This led to distributed storage and processing systems running on commodity hardware flourishing since the early 2000s, with great influence from Google's GFS \cite{gfs} and MapReduce papers, which stood at the foundation of the Apache Hadoop ecosystem. Since then, dozens of products have emerged in this field, improved and adapted for different needs. 

One of such needs is reliably processing large streams of data in real-time, which is especially crucial considering the vast amount of internet traffic nowadays \cite{streaminganalytics}. For example, a video streaming service could want to analyze which videos are most popular in specific regions at any given moment so that they could automatically cache them on close by servers for better loading times.

Even though there are quite a few existing distributed streaming processing systems, they usually have their own caveats and inefficiencies, which will be discussed in further chapters. It must be also noted that integrating an open-source solution into an existing infrastructure is often problematic and ineffective. 

Devising an efficient streaming processing algorithm and building a flexible system around it could benefit a lot of internal teams at Yandex and improve its services, which is directly related to expenses and revenue. Meanwhile, the need to tolerate inevitable machine failures in large clusters and to provide various consistency guarantees ads algorithmic complexity and academic merit to this work.

\subsection{Problem overview, goals and achieved results}
\label{subsection:pgr}
\begin{figure}[htbp]
  \centering
  \includegraphics[scale=0.66]{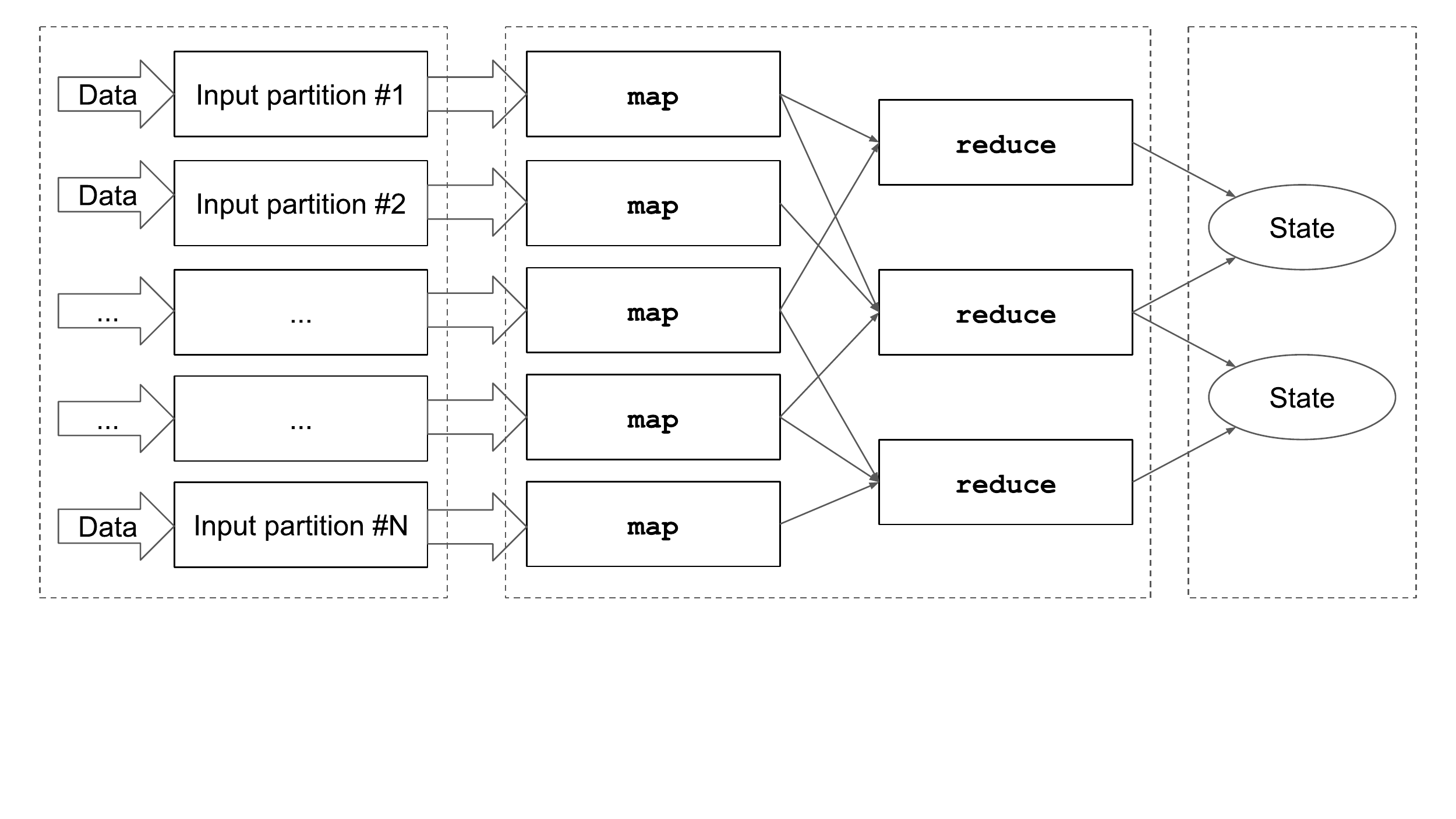}
  \vspace{-2.3cm}
  \caption{Computation model.}
  \label{fig:processing-model}
\end{figure}

In the realms of this thesis we will work in the computation model described in the paragraphs below and displayed in figure \ref{fig:processing-model}.

The input is given as a stream of rows consisting of multiple partitions. Similar to Kafka \cite{kafka}, each of the partitions constitutes a queue of rows. Producers can append rows to the end of these queues and consumers can read the partitions at their own pace. There are multiple services used throughout Yandex that efficiently support this interface. 

As in classic MapReduce, each of the partitions is read by a mapper worker executing a user-provided \verb|map| function. For each produced row an index of the reducer worker that is supposed to handle it can be computed. We call this the \textit{shuffle function}, which is required to be deterministic. Reducers execute a user-provided \verb|reduce| function, which will typically interact transactionally with some reliably stored internal state.

Even though there are more expressive models and systems supporting large computation graphs, designing such a system was out of scope of this project. Conversely, we focused on an efficient and flexible implementation of the aforementioned \verb|map|-\verb|reduce| stage, usually called the \textit{shuffle stage}, which is often a weak point of the larger products. More specifically, we worked to reduce \textit{write amplification} and devise a solution with a low overhead on persistent storage, which is a somewhat novel approach in this field.

Thus, the aim of this project was to design and implement the underlying infrastructure for a system operating within the interface described above and satisfying the following requirements:  
\begin{enumerate}[rightmargin=\parindent]
    \item Exactly-once semantics: the effect of processing each row should only be observed once, as part of a successful transaction commit to the reducer's internal state.
    \item Fault-tolerance in regard to failures of any workers in the system.
    \item The ability of healthy reducers to continue working successfully amidst failures of others.
    \item The ability of the system to continue working successfully amidst slowdowns and failures of individual partitions.
    \item General CPU and memory efficiency, including low write amplification factors.
\end{enumerate}

As a result, we have implemented a solution fulfills all of the above conditions, except being able to tolerate overly lengthy downtimes of reducers. Even so, we propose a design that would rectify this problem, along with other potential enhancements, in chapter \ref{section:fw}. Our system can process gigabytes of streaming data per second and perform real-time analysis on it with sub-second latencies.

\section{Related work}
\label{section:rw}
As mentioned in the previous chapter, the work on real-time data processing systems has been very fruitful in the last ten years. Many open-source and commercial solutions are available on the market. Some of them, either prominent or specifically related to our planned approach, are described in more detail in the subsections below.

Even though our approach is arguably quite novel, it is important to remember that due to the tremendous overheads and limitations of applying an existing open-source solution at a company as large as Yandex some similarities with existing systems are expected and even beneficial. We try to leverage the best ideas from state of the art systems with our own propositions and Yandex's unique infrastructure.

\subsection{MapReduce}
Google's MapReduce and its open-source implementation within Apache Hadoop are in no way equipped to handle streaming processing. However, their approach is at the foundation of our system and many others, so it is important to provide an overview of the paradigm and highlight its limitations. 

MapReduce handles input in the form of key-value pairs stored in a distributed file system as a large number of small-sized splits. During the \textit{map} phase, workers each read a designated number of splits and execute a user-defined function on the collected key-value pairs. The produced results are partitioned by key and stored on local disks by the mappers. Each partition is assigned to a separate specific worker. These workers collect their corresponding partitions during the \textit{reduce} phase, combine and sort them by key, and execute another user-provided function on the result. The output is written to a file in the distributed file system. It is ensured by the partitioning that pairs with the same key are handled by the same reducer. In case of failures, workers can be simply restarted.

As with many other applications, this approach is heavily bounded by the latencies of I/O operations, such as reading and writing files to and from the distributed file system and local disks. It is also not easy to provide any reasonable consistency guarantees if reducers were to modify a global state or try to produce partial outputs. Thus, we take a different approach for delivering data between phases to accommodate for streaming processing needs, only borrowing the general model, which has proven to be adaptable to an abundance of diverse tasks.

\subsection{MapReduce Online}
MapReduce Online \cite{mronline} is an attempt to specifically tackle the issues mentioned above to improve and expand the abilities of Hadoop MapReduce, including the option to run \textit{continuous} jobs on a stream of data. 

The main enhancement comes in the form of pipelining. Instead of always writing data to disk, mappers now collect batches of key-value pairs and send them to the appropriate reducers before the whole input is mapped. To preserve fault-tolerance guarantees, these batches are still written to storage. However, they are typically retrieved soon after that, when they are presumably still resident in cache. Combined with periodic writes of reducer outputs to HDFS, an open-source implementation of the Google File System, this approach can be used to process data in a streaming fashion.

Even so, the solution in question has many faults. It still persists data during both MapReduce phases, incurring high factors of write amplification. Moreover, it is not equipped to handle straggling reducers, which would cause unsent batches to build up, hindering the benefits of the approach above. It is also important to note that the delivery semantics of the proposed solution are unclear, and it does not seem to guarantee exactly-once processing without manual handling.

We aim to build on the general idea of sending reasonably small batches of data from mappers to reducers as soon as possible and provide better write amplification factors and stronger processing guarantees.

\subsection{Apache Spark Streaming}
Spark Streaming \cite{sparkstreaming} proposes the notion of \textit{discretized streams}, which structure streaming computations as a series of stateless deterministic batch computations. The underlying units of these computations are stored in Resilient Distributed Datasets \cite{rdd}, a data structure already utilized in regular Apache Spark \cite{spark}. RDDs are kept in memory and achieve fault-tolerance by storing the sequence of transformations that needs to be applied to the original data in order for the RDDs to reach their current state. In case of failure, these computations can simply be replayed. RDDs are internally stored as multiple partitions, which allows to perform data transformations and recovery in parallel across multiple nodes. Moreover, recovery can also be conducted independently across different RDDs.

Spark Streaming supports \textit{map} and \textit{reduce} transformations among others and is well integrated with regular Spark batch processing. However, when performing reduce-like operations Spark employs a shuffle algorithm similar to the classic MapReduce implementation, collecting inputs for reducer tasks on disk, which is a big overhead and known weak point of the system. Additionally, while it is possible to achieve end-to-end exactly-once guarantees with Spark Streaming when using the Kafka Direct API for input, it requires manually implementing transactional outputs.

While designing as large a system is out of the scope of this project, we capitalize on the idea of only persisting computation \textit{meta-state} and keeping most of the handled data in memory, even during the shuffle operation. We also provide more out-of-the-box options for end-to-end exactly-once processing and atomic interactions with consecutive batches of data, which are heavily limited in Spark Streaming.

\subsection{Apache Storm}
Storm \cite{storm} is a distributed fault-tolerant real-time data processing system that was developed and open-sourced by Twitter. It allows to process streams of tuples flowing through computational graphs, called \textit{topologies}. Topology nodes are subdivided into two categories: \textit{spouts}, which represent data sources, and \textit{bolts}, which represent processing operators. Storm supports parallelism in both of these kinds of tasks and is able to read from partitioned queues, such as Apache Kafka.

Storm bolts can perform a variety of grouping operations, which essentially perform a shuffle and send data to different receiving instances of the same bolt. Internally, in-flight tuples are stored using in-memory queues and only sent to their corresponding receivers over the network without being persisted to disk, similar to the approach we take in our system. Storm seems to store its persistent state using Apache ZooKeeper \cite{zk}, even though it is not exactly clear from the paper cited above how exactly it is used to guarantee fault-tolerance. However, the biggest disadvantage of Storm comes with its weak message processing guarantees, only supporting either at-least-once or at-most-once semantics. Moreover, the mechanism to achieve at-least-once delivery is quite complex, implemented by tracking linage of each tuple and requiring the user to manually add these dependencies for a tuple and send acknowledgement events once it is processed.

We use a similar tactic for performing shuffle operations without unnecessary write overheads, at the same time employing a novel approach for providing strong exactly-once processing guarantees without any additional implementation hassles for our system's clients.

\subsection{Apache Flink}
Flink \cite{flink} is a distributed fault-tolerant streaming and batch data processing system, with its creators and many committers employed at Ververica, formerly called data Artisans. It is one of the few systems, along with Spark Streaming, to provide a specialized API for both variations of data processing within one all-encompassing system. Internally, Flink express pipelines as dataflow graphs, consisting of sources, sinks and potentially stateful data operators, as well as data stream nodes that represent records produced by an operator or input source and available for consumption by other operators. Parallel execution is performed by splitting streams into multiple partitions and executing operators on each of them in different concurrent subtasks. This underlying representation is shared by both the batch DataSet API and the streaming DataStream API. As with many of the systems described, durable message queues like Apache Kafka are prominent and popular data sources for Flink.

Flink provides several common processing abstractions that require the inputting data stream node to perform a shuffling operation between different partitions. In the case of batch processing, the shuffle is performed in a classical MapReduce fashion, with intermediate data designated for a certain partition being persisted to disk. For streaming processing, however, Flink utilizes a transient network-only shuffle approach. To guarantee fault-tolerance a unique method of checkpointing operator state to persistent storage like Apache Hadoop HDFS is implemented, called Asynchronous Barrier Snapshotting \cite{abs}. The novelty lies in the fact that for acyclic execution graphs no actual data records have to be stored in the persisted checkpoint. Flink data sources regularly insert special control barrier-records into streams, which trigger an asynchronous snapshotting operation when encountered by executing nodes. The system is then able to restart from these snapshots in case of failures and replay the processing with exactly-once guarantees.

Our solution shares a lot of similarities with Flink, providing an efficient network-shuffle implementation that is also fault-tolerant and only commits external state updates exactly once per record. However, we take a somewhat different approach to achieving this, which makes our persisted state more compact, especially in cases of potential windowed aggregation where Flink does end up storing in-flight records in the checkpointed state. We also utilize YT's own persistent store, which is more robust than HDFS. Additionally, our system provides a wider range of transactional interactions with the output state, whereas Flink can only guarantee that effects will only be applied once if a Kafka transactional producer is used as a sink.

\section{About YT}
\label{section:yt}
The resulting effort is part of Yandex YT, which is the main driver behind all kinds of distributed systems at Yandex. In order to achieve the desired fault-tolerance and delivery semantics outlined in subsection \ref{subsection:pgr}, our algorithm takes advantage of a few other services offered by the YT ecosystem. In this subsection we provide some details about these products, which are necessary for understanding the proposed design. 

Most important for our case are YT's dynamic tables. They are architecturally similar to BigTable \cite{bigtable} and HBase \cite{hbase} and guarantee fault-tolerance and consistency using Hydra, Yandex's original consensus protocol similar to Raft \cite{raft}. There are two types of dynamic tables offering different functionalities and interfaces. Ordered tables behave similarly to Kafka's topics, which were touched upon in subsection \ref{subsection:pgr}. These tables will be covered in more details in later subsections. Sorted tables provide a typical row-based strictly schematized storage supporting fine-grained reads and writes. Users can interact with these tables atomically by creating transactions, which can span across multiple rows and both kinds of tables. Transactions are implemented using two-phase commits, similar to the approach in Google's Spanner \cite{spanner}.

Additionally, we use Cypress, a filesystem-like metainformation store, which can also keep an attribute mapping in its nodes and supports transactions and locks. This allows it to be used similarly to Apache ZooKeeper. Internally it utilizes the same Hydra consensus algorithm mentioned above.

\section{System architecture}
\label{section:sa}
In very general terms, a single streaming task, which we call a \textit{streaming processor}, consists of endlessly running mapper and reducer jobs. Mappers read their corresponding partitions and keep a rolling window of \verb|map|ped rows in memory. These rows are split into small batches and re-read and re-mapped in case of failure. Mappers also compute the shuffle function for every row and store the results alongside these batches. Reducers, in turn, pull the corresponding rows from the mappers and process these rows using the specified \verb|reduce| function. The user-provided code can open a transaction while processing a batch of rows and modify a dynamic table of its choice. The system will then commit the required internal meta-state changes in the same transaction, guaranteeing that the effect of processing a batch of rows is applied exactly once. Mappers update their own persistent meta-state and move their window forward once rows are successfully processed by reducers.

In the following subsections we will delve deeper into the design and components of the proposed solution.

\subsection{User API}
\label{subsection:user_api}
The whole system operates within a schematized key-value row-based data model, encapsulated in the \verb|UnversionedRow| class. It is stored as an array of strictly-typed data values, with a separate \verb|NameTable| object used to map the array's indexes to the corresponding key strings. An \verb|UnversionedRowset| object stores an array of \verb|UnversionedRow| objects along with a \verb|NameTable| instance. This is the main abstraction users can interact with.

To run their own streaming processor, users have to provide C++ implementations of the two following interfaces.

\newpage
\subsubsection{Mapper}
\begin{lstlisting}
struct PartitionedRowset {
    UnversionedRowsetPtr Rowset;
    std::vector<int> PartitionIndexes;
};

class IMapper {
public:
    virtual PartitionedRowset Map(UnversionedRowsetPtr rows) = 0;
};

IMapperPtr CreateMapper(INodePtr configNode, IClientPtr client, TableSchemaPtr schema, MapperSpecPtr spec);
\end{lstlisting}

The \verb|Map| function receives a batch of rows and has to return a new, possibly empty, \verb|UnversionedRowset| object along with a vector of the same size, indicating to which reducer each produced row should be sent to. The returned batch can have a different schema and contain more or fewer rows than the input. In other words, it represents a one-to-many mapping for each single input row. The \verb|Map| function must be deterministic, otherwise exactly-once processing cannot be guaranteed. The \verb|CreateMapper| function should create an instance of the user's derived class, given the user's own specified configuration (see subsection \ref{subsection:configuration}), a YT client that can be used to perform operations with other YT components, the schema of the input, as well as the specification of this mapper within the streaming processor. The last of these parameters contains the number of reducers, which is often useful for the user's partitioning logic.
\newpage
\subsubsection{Reducer}
\begin{lstlisting}
class IReducer {
public:
    virtual ITransactionPtr Reduce(UnversionedRowsetPtr rowset) = 0;
};

IReducerPtr CreateReducer(INodePtr configNode, IClientPtr client, ReducerSpecPtr spec);
\end{lstlisting}

The \verb|Reducer| function receives a batch of rows designated to this reducer and can perform arbitrary user-defined processing. If the processing includes modifications of YT tables, the user can start a new transaction, act upon some table rows and return this transaction without committing. The reducer instance will modify the internal meta-state in the returned transaction and then try to commit both changesets atomically, guaranteeing exactly-once semantics for reducer processing. The user may also return \verb|nullptr|, in which case the reducer instance will start a transaction itself. More details on this will follow in subsections \ref{subsection:reducer_workflow} and \ref{subsection:fault_tolerance}. The \verb|CreateReducer| function should create an instance of the user's derived class, given the user's own specified configuration (see subsection \ref{subsection:configuration}), a YT client that can be used to start transactions and perform operations with other YT components, as well as the specification of this mapper within the streaming processor.

\subsection{Input model}
\label{subsection:input_model}
As mentioned earlier, our system accepts inputs presented in a Kafka-like fashion as multiple queues (partitions) organized into one stream (topic), which are assumed to be stored reliably.

To be more exact, a viable input source needs to implement two methods of  the \verb|IPartitionReader| interface, objects of which are responsible for retrieving the data from a single input partition.

The first method, \verb|Read|, takes two integers \verb|beginRowIndex|, \verb|endRowIndex| and a more freely defined \verb|continuationToken| as parameters and should return the next batch of rows, along with a continuation token pointing to the next position in the stream. The \verb|continuationToken| parameter indicates a starting position in the input partition, from which this batch of rows should be read. The returned rows will be assigned indexes starting with \verb|beginRowIndex| in the corresponding mapper's \textit{input numbering}. Thus it is essential that this method returns rows in a deterministic order, otherwise there is no way for our system to guarantee any reasonable delivery semantics. The \verb|endRowIndex| parameter serves as a hint on the number of rows to read. 

The second method, \verb|Trim|, takes an integer \verb|rowIndex| and the aforementioned \verb|continuationToken| as parameters and should, in some way, mark entries before the \verb|continuationToken| or with index lower than \verb|rowIndex| as committed and thus safe to delete. Logically, this method must be idempotent. It is also allowed to perform this action asynchronously at some later time.

The \verb|continuationToken| can be of any serializable type specific to the input source, however, it must be noted that it will be stored within the mapper's persistent state.

Currently, our system supports the following two internal data delivery services:
\begin{itemize}[rightmargin=\parindent]
    \item Reading from an ordered dynamic table. It is internally divided into queue-like partitions called tablets. Each tablet is indexed from zero in an absolute fashion and can be read from and trimmed using these indexes. This makes it easy to use with the \verb|...Index| arguments in the methods above.
    \item Reading from a LogBroker topic. It is internally divided into partitions. These partitions have their own offsets,  which increase monotonically, but are not guaranteed to be sequential. Thus, it is necessary to use the \verb|continuationToken| argument to specify the next offset to read from in each cluster. 
\end{itemize}

The complexity of this design is largely due to the order in which the input sources listed above were supported. It will be somewhat remodeled and simplified in the future.

\subsection{Mapper workflow}
Below we will lay out the runtime of a single mapper instance.
\subsubsection{Internal state}
\label{subsubsection:internal_state}
Mappers maintain two kind of absolute numberings that increase sequentially as the streaming processor is working:
\begin{itemize}[rightmargin=\parindent]
    \item The \textit{input numbering} pertains to rows read by the mapper's partition reader instance.
    \item The \textit{shuffle numbering} pertains to rows produced by the user-provided \verb|Map| function applied to the rows above.
\end{itemize}

The following entities stored by a mapper instance are vital to the execution of our proposed algorithm:
\begin{itemize}[rightmargin=\parindent]
    \item An instance of the \verb|IPartitionReader| interface, described in subsection \ref{subsection:input_model}. It encapsulates all interactions with the input stream.
    \item A queue of \verb|WindowEntry| objects, which hold information about batches of read and mapped rows. These entries are indexed sequentially within the lifetime of the instance, thus we also store the absolute index of the first entry in the queue.
    \item An array of \verb|BucketState| objects, one for every reducer, which hold a queue of shuffle row indexes that will need to be shipped to said reducer, along with the window entry index in which the first of these rows is to be found.
    \item \verb|LocalMapperState|, a local copy of the persistent state (see \ref{subsubsection:persistent_state}), which serves as a lower bound on yet unread row indexes.
    \item \verb|PersistedMapperState|, the current version of the persistent state as seen by this mapper.
\end{itemize}
Each window entry also stores a \textit{bucket pointer count}, which tallies the number of buckets for which this entry holds the first row in their queue.
\subsubsection{Persistent state}
\label{subsubsection:persistent_state}
The persistent state is stored in a sorted dynamic table shared by all mapper instances. Mappers are indexed starting from 0, and every mapper knows its index from its configuration, which will be later discussed in subsection \ref{subsection:configuration}. Each mapper only interacts with its single corresponding row of the table and doesn't interfere with other running mappers. The state table contains the following columns:
\begin{itemize}[rightmargin=\parindent]
    \item \verb|mapper_index|: the key column.
    \item \verb|input_unread_row_index|: the index in the input numbering of the first row that was not yet successfully processed and committed by its corresponding reducer.
    \item \verb|shuffle_unread_row_index|: same as above, but in regards to the shuffle numbering.
    \item \verb|continuation_token|: same as above, but in terms of the partition reader's continuation token.
\end{itemize}

This state is used to guarantee consistency and exactly-once semantics in case of failures, which will be discussed in more details in subsection \ref{subsection:fault_tolerance}.

\subsubsection{Input ingestion procedure}
\label{subsubsection:mii}
This procedure starts as soon as the mapper is alive. Initially, it fetches its corresponding row from the state table. As mentioned in \ref{subsubsection:internal_state}, a local copy is stored in both the \verb|LocalMapperState| and \verb|PersistedMapperState| field.

This state is then read into variables \verb|inputCurrentRowIndex|,\\\verb|shuffleCurrentRowIndex| and \verb|continuationToken|. 

Afterwards, the following cycle is repeated continuously while the instance is working:
\begin{enumerate}[rightmargin=\parindent]
    \item Wait for a configuration-defined amount of time if the previous iteration of this cycle didn't finish with appending a non-empty batch of rows to the internal state.
    \item Wait for the next batch of rows from the mapper's partition reader instance. 
    \item Fetch the current remote persistent state, skipping to the next iteration in case of errors. If the result differs from the state stored in \verb|PersistedMapperState|, we are in a split-brain situation and the mapper waits out a configurable delay, after which the internal state is dropped and the whole input ingestion procedure described here in \ref{subsubsection:mii} is restarted.
    \item If the batch is empty, skip to the next iteration of the cycle. Otherwise, the resulting rows now have sequential indexes starting with \verb|inputUnreadRowIndex|.
    \item Run the user-provided \verb|Map| function on the batch of rows and build a \verb|WindowEntry| instance, which contains the returned rowset, the corresponding index ranges in both numberings, the continuation token returned from the partition reader, as well as additional information mentioned in \ref{subsubsection:internal_state}.
    \item Increment the memory usage semaphore and push the built window entry to the mapper's \verb|WindowQueue|. Iterate over the mapped rows and push their (shuffle) indexes to the corresponding reducer buckets, incrementing the entry's bucket pointer count when adding the first element to a bucket. In the latter case, the bucket's first window entry index is set to the index of the current entry.
    \item Update the \verb|...CurrentRowIndex| and \verb|continuationToken| variables accordingly.
    \item If the memory limit is exceeded, block on the semaphore, waiting for the usage to be below the threshold again.
\end{enumerate}

\subsubsection{RPC methods}
Concurrently with the actions described above, each mapper responds to \verb|GetRows| remote procedure calls from reducers. This method is described by the following protobuf-schema:
\vspace{0.5\baselineskip}
\begin{lstlisting}
message TReqGetRows {
    optional int64 count = 1;
    optional int64 reducer_index = 2;
    optional int64 committed_row_index = 3;
    optional string mapper_id = 4;
}
message TRspGetRows {
    optional int64 row_count = 1;
    optional int64 last_shuffle_row_index = 2;
}
\end{lstlisting}

With this call the reducer with index \verb|reducer_index| requests \verb|count| of its assigned rows. The \verb|committed_row_index| parameter denotes the shuffle index of the last row successfully processed and committed by this reducer. The \verb|mapper_id| parameter is used to discard incorrect requests due to stale discovery information (see subsection \ref{subsection:configuration}). The actual rows are returned as attachments in a binary format.

As a result, \verb|row_count| rows are returned, with \verb|last_shuffle_row_index| indicating the shuffle index of the last of these rows. The latter of these response fields is needed due to the fact that rows assigned to each reducer don't necessarily have sequential indexes. The amount of returned rows can be smaller than the requested number or even zero.

The execution of this procedure by the mapper boils down to the following steps:
\begin{enumerate}[rightmargin=\parindent]
    \item If \verb|mapper_id| differs from the mapper's own GUID return an error response.
    \item Pop from the reducer's corresponding \verb|BucketState| while the first index in the queue is less than or equal to \verb|committed_row_index|. Iterate across the beginning of the window entry queue to update the bucket's first window entry index and the windows' bucket pointer counts as necessary.
    \item Schedule trimming operations if necessary, see \ref{subsubsection:trimming}. 
    \item Serialize \verb|count| rows, or as many as are available, from the beginning of the bucket's queue and return them as attachments along with the appropriate response fields. It is important to note that these rows are not deleted from the queue.
\end{enumerate}

\subsubsection{Trimming}
\label{subsubsection:trimming}
In order to make progress a mapper needs to free up memory used by rows that were successfully processed by their corresponding reducers. This is done by the \verb|TrimWindowEntries| method, which pops window entries with bucket pointer counts equal to zero from the front of the window queue and increments the absolute index of the first window in the queue accordingly. The after-the-end indexes and continuation token of the last popped window entry, if there was one, are used to update the \verb|LocalMapperState| field.

To support end-to-end exactly-once scenarios a mapper has to manually ensure that its input partition moves along as rows are processed by reducers, which also requires updating its persistent state so that the mapper doesn't try to read already-deleted rows. Timely updating persistent state is also necessary to reduce the number of already-processed rows that will be reread by the mapper when it restarts after a possible failure. This is implemented by the \verb|TrimInputRows| method. It opens a dynamic table transaction and fetches the current committed persistent state in it. If it is equal to the state stored in \verb|PersistedMapperState| and \verb|LocalMapperState| is further along than the committed state, the method tries to update the remote state with \verb|LocalMapperState| within the same transaction. If the transaction was committed successfully, the method updates \verb|PersistedMapperState| with the committed result and calls \verb|Trim| on the partition reader (see subsection \ref{subsection:input_model}), passing the input index and continuation token from the local state as arguments. 

By using the \verb|LocalMapperState| field we were able to separate trimming actions into two methods that can be executed independently, which allows for a more efficient asynchronous implementation. We call the first method when a \verb|GetRows| call causes a bucket pointer count to become zero. We schedule the second method, which is more costly due to its transactional interactions with dynamic tables, to be called regularly with a configuration-defined period, usually on the order of a few seconds.

\subsection{Reducer workflow}
\label{subsection:reducer_workflow}
Below we will describe the runtime of a single reducer instance, which is arguably more straightforward than the life of the mappers.
\subsubsection{Persistent state}
The persistent state is stored in a sorted dynamic table shared by all reducer instances. Reducers are indexed starting from 0, and every reducer knows its index from its configuration, which will be later discussed in subsection \ref{subsection:configuration}. Each reducer only interacts with its single corresponding row of the table and doesn't interfere with other running reducers. The state table contains the following columns:
\begin{itemize}
    \item \verb|reducer_index|: the key column.
    \item \verb|committed_row_indices|: a list of shuffle row indices, one for each mapper, indicating that all rows up to said index were reliably processed by the reducer.
\end{itemize}

This state is used to guarantee consistency and exactly-once semantics in case of failures, which will be discussed in more details in subsection \ref{subsection:fault_tolerance}.
\subsubsection{Main procedure}
\label{subsubsection:reducer_main_procedure}
This procedure starts as soon as the reducer is alive, and performs the following cycle continuously while the instance is working:
\begin{enumerate}
    \item Wait for a configuration-defined amount of time if the previous iteration of this cycle didn't finish with a successful persistent state update.
    \item Fetch the current persistent state into \verb|reducerState|.
    \item Fetch a list of mappers from discovery (see subsection \ref{subsection:configuration}). Build asynchronous RPC \verb|GetRows| requests to these mappers, one per mapper-index, passing the corresponding value from \verb|reducerState| as \verb|committed_row_index|. Wait for all of these requests to complete. Only one request per mapper index is made.
    \item Create \verb|newReducerState| as a copy of \verb|reducerState| with each array element set to the \verb|last_shuffle_row_index| value returned by the corresponding mapper. If a mapper with a certain index returned an empty batch of rows, an error or was missing in discovery and wasn't polled, its entry is left unchanged. If all \verb|row_count| variables are equal zero the next steps are skipped.
    \item Deserialize the attachments in RPC responses to the requests above into rows and run the user-provided \verb|Reduce| function on all of these rows combined into one batch.
    \item If the \verb|Reduce| call returned null (see subsection \ref{subsection:user_api}), start a new transaction to commit updates to the persistent state. Otherwise, the actions in the following steps are performed within the transaction returned by \verb|Reduce|.
    \item Fetch the persistent state again within the transaction. If it differs from \verb|reducerState|, we are in a split-brain situation and skip to the next iteration of the cycle without committing anything.
    \item Write \verb|newReducerState| to the reducer's corresponding state row and try to commit the resulting transaction.
\end{enumerate}
Some of these steps can produce errors, such as a failed state fetch or transaction commit. If that happens, we just skip forward to the next iteration and wait out the back-off delay in step 1.

\subsection{Configuration, discovery and control}
\label{subsection:configuration}
The system is configured using YT's own JSON-like format, called YSON. There are quite a few parameters of the algorithm described above which can be tweaked by the user, which, however, are too minor to be discussed in this paper. Additionally, users can define their own similar configuration classes, which they can use to specify parameters for their own \verb|Map| and \verb|Reduce| implementations. Each mapper and reducer is also passed a separate system-generated specification file which contains the GUID of the streaming processor, the path of the corresponding state table, the worker's index and GUID, as well as the number of reducers or mappers respectively.

We utilize an existing YT component for performing discovery, which is required for reducers to be able to resolve the mappers' addresses.  Internally, it uses Cypress, described in chapter \ref{section:yt}. Participants of a discovery group create and take a lock on key-named nodes in a shared Cypress directory, storing any necessary information in the node's attributes. The directory's name, therefore, represents the name of this discovery group. Other clients can fetch a list of nodes in this directory and retrieve the relevant attributes.

Mappers all join the same discovery group, providing their GUID's as keys, and store their address, RPC port and index as node attributes. Reducers also join a similar group, sharing only their index. It is important to understand that in case of failures, or even on startup, the information in these discovery groups can be stale and taking some time to update. For example, a failed mapper and it's newly-alive replacement could temporarily both appear in discovery. Thus,  we have to take the additional precautions in the main reducer procedure which were described in \ref{subsubsection:reducer_main_procedure}.

The whole streaming processor is executed as a YT ``vanilla'' operation, which allows running user-specified binaries on a number of nodes, automatically restarting them in case of failures. Currently, there is a manual script that sets up such an operation given the appropriate configuration files. In the future, however, a controller currently under development will be used to start and monitor a streaming processor, correctly restarting the whole operation if it spuriously fails.

\subsection{Fault tolerance and exactly-once delivery}
\label{subsection:fault_tolerance}
In the examination of our system we consider that any worker can fail spontaneously. Moreover, since workers failed workers are automatically restarted (see subsection \ref{subsection:configuration}) we can temporarily end up with multiple instances of the same mapper or reducer if network partitions occur, producing a so-called split-brain scenario.

The proposed solution can maintain correctness and exactly-once delivery semantics in the scenarios described above. This is guaranteed by the following simple conclusions:
\begin{itemize}
    \item A mapper's state is only advanced past a row once all of the rows that were produced from it by the \verb|Map| function have been successfully processed by their designated reducers.
    \item An input row is only trimmed once the corresponding mapper's state has advanced past it. Thus, rows will not be trimmed unless they were at least once.
    \item A produced row is only sent to its designated reducer if the corresponding mapper's state was not modified by some other worker while the originating input row was being read. This, along with the \verb|Map| function being deterministic, ensures that rows receive the same input and shuffle indices even in split-brain scenarios. 
    \item Reducers always process rows and modify their persistent state atomically, thus a row is guaranteed to be processed at most once even in split-brain scenarios.
\end{itemize}
Even though we mostly talk about split-brain scenarios in the points above, simple failures are essentially a subset of these scenarios and usually more straightforward. If a mapper or reducer is restarted it looses some progress, but correctness is maintained for the same reasons as mentioned above.

Besides being able to handle failures correctly, another important aspect is that healthy workers still achieve progress even when others have failed. In our current solution this is true for mappers: a failed or unavailable mapper or input partition is simply ignored by reducers until it comes back online, so the streaming processor is not hindered at all. When reducers fail, however, the mappers' won't be able to trim rows designated to this reducer, causing their row windows to build up and hit the memory limit. This would eventually cause the whole processor to stagnate. An improvement able to overcome this problem has already been designed and is described in chapter \ref{section:fw}.

\section{Evaluation}
\label{section:evaluation}
As a result of this thesis we implemented the design described above within YT. The core of the system itself was written in C++, and Python was used for tests. The current implementation contains more than 6000 lines of code, about 4000 of which are in C++. In the subsections below we describe how we have tested our solution to assess its correctness, performance and practicality.

\subsection{Local integration tests}
We check the system's correct behaviour in various testing scenarios by using a Python environment that sets up its own small local YT cluster and LogBroker installation. To allow for more intricate checks, we implemented mappers and reducers that interpret control strings within the stream being processed and either halt their execution for a specified amount of time or use Cypress nodes to halt and wait for an external signal to continue. Altogether, this allowed us to write integration tests that verify the correctness of the intermediate state during and after simulated job failures and automatic restarts.

\subsection{Performance testing}
To analyze performance we deployed a streaming processor on a production-like testing cluster to perform somewhat realistic analysis on real-time high-velocity stream of data. We chose to experiment with LogBroker as an input provider since it is the more widespread data delivery solution at Yandex.

Our input source was a topic fed by logs from YT's replicated master nodes, which are responsible for almost all crucial internal YT coordination within a single cluster. The topic in question has 90 partitions, each actually representing 5 distinct partitions across different clusters, which makes a total of 450 independent unique input partitions. The write rate to the topic is steady at around 3.5 gigabytes of uncompressed data per second, with messages consisting of batched and joined master node log entries. Since every source partition has to be read by a designated mapper, our processor consisted of 450 mapper jobs which were pulling decompressed messages from the LogBroker topic. The mappers' \verb|Map| implementation split each read message back into individual log messages. These messages were then parsed and hash partitioned by their respective \textit{user} and \textit{cluster} fields. Log messages that didn't have a \textit{user} field were simply ignored, which eliminated around 80 to 90 percent of all messages. The remainder was processed by 10 reducer workers, which grouped messages by user and cluster, writing the timestamp of the user's last access to the cluster and a tally of the number of corresponding messages in the batch to a sorted dynamic table shared by all reducers. Since many of the incoming messages were dropped by mappers, the resulting input flow into the reducers was around 400 megabytes per second. 

This last number might not seem too big, but it must be noted that our setup shares a lot of characteristics and caveats with real-life processing tasks: the write rate into individual partitions varies with time and even more across clusters, mappers perform significant filtering work and the distribution of keys is uneven, with \verb|root| and a few other system users appearing in overwhelmingly more messages than regular users.

\begin{figure}[H]
\includegraphics[width=\textwidth]{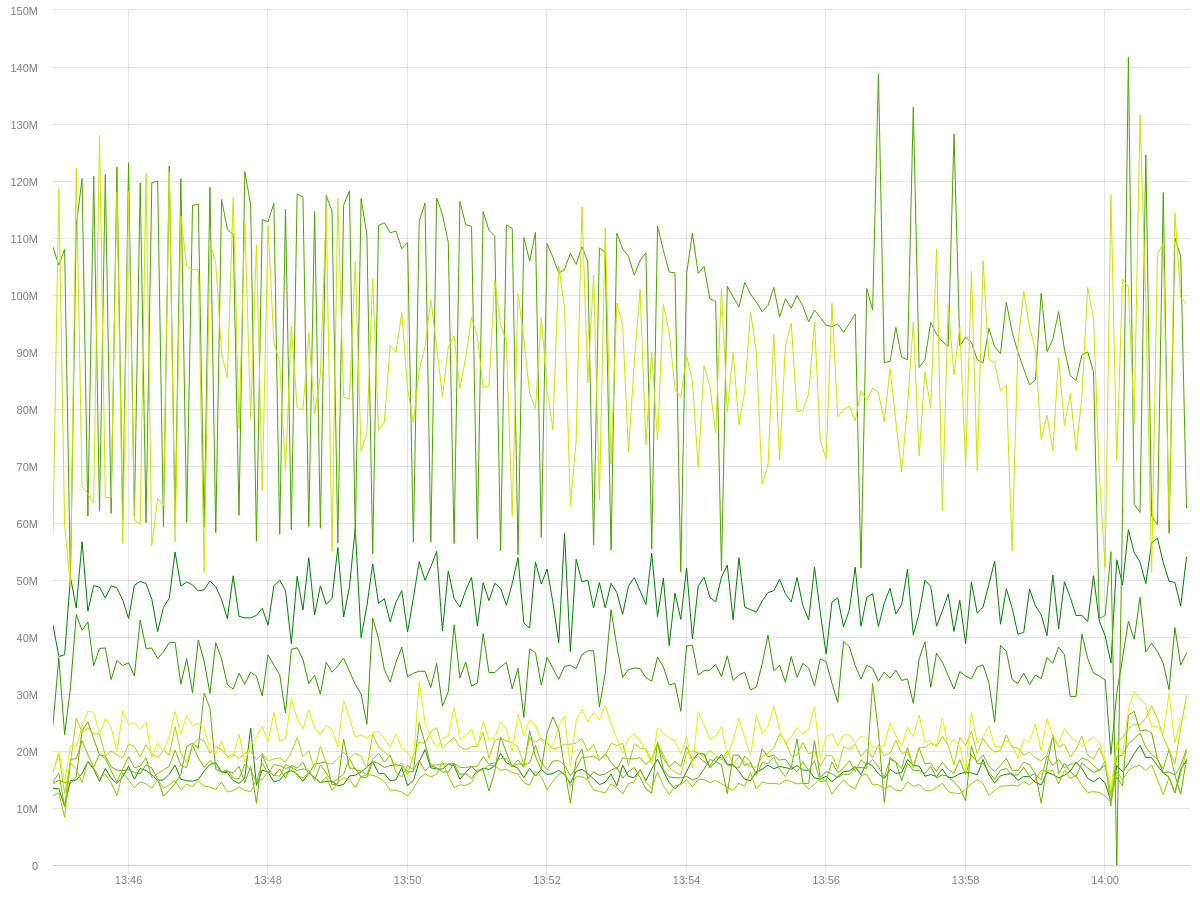}
\centering
\vspace{-1cm}
\caption{Reducer throughput.}
\label{fig:reducer_throughput}
\end{figure}

\begin{figure}[H]
\includegraphics[width=\textwidth]{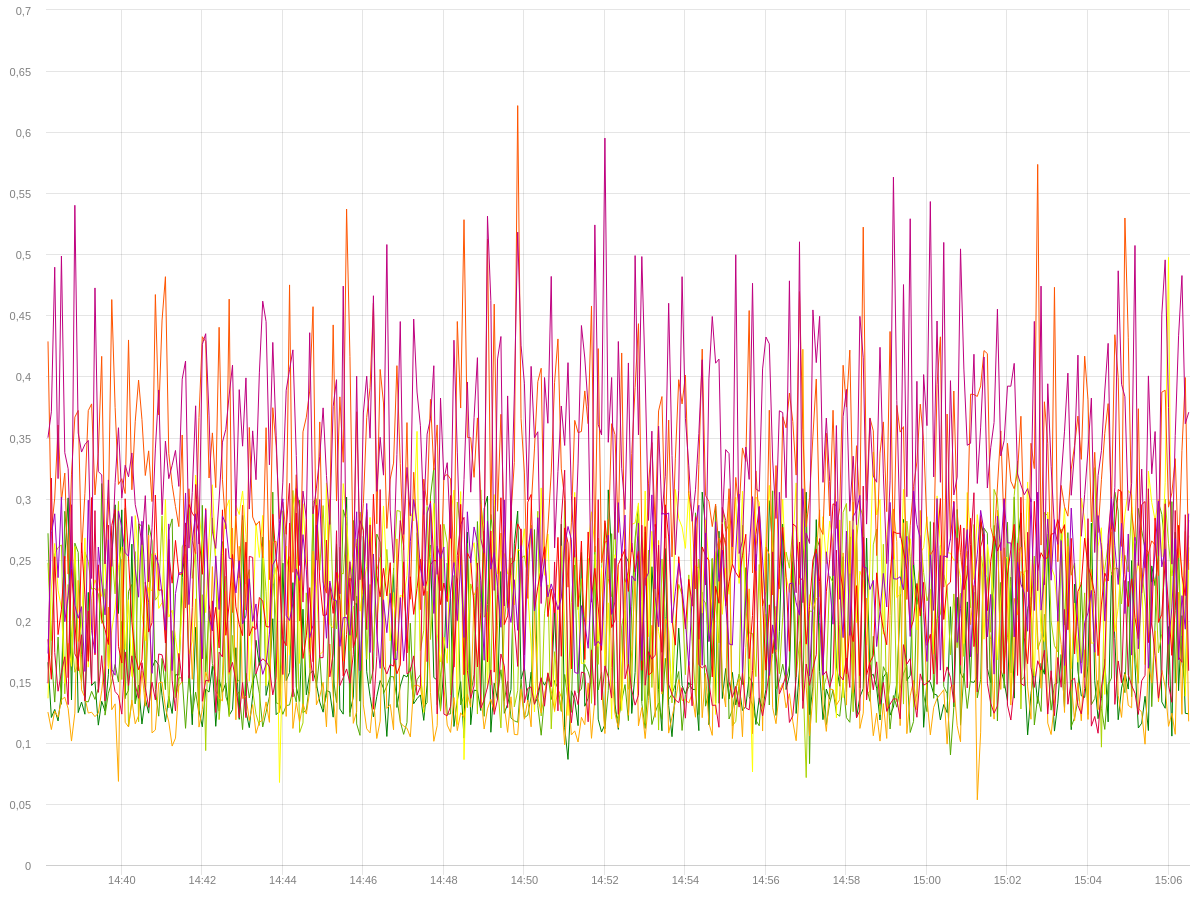}
\centering
\vspace{-1cm}
\caption{Read lag from 10 selected mappers.}
\label{fig:read_lag}
\end{figure}

As can be seen in figure \ref{fig:reducer_throughput} above, reducers are able to process up to around 95 megabytes per second. The maximum input ingestion speed by reducers is the relevant metric here since we know that the data is quite uneven, causing the most loaded reducers to become bottlenecks for the whole processor. Another important metric to watch is the read lag, defined as the elapsed time between a message being written to the LogBroker topic and the moment when the message was read by the corresponding mapper. In our experiment mappers were able to work with a steady read lag of a few hundred milliseconds, which can be seen for ten mappers in figure \ref{fig:read_lag} on the next page. We chose these mappers evenly across partitions from different clusters since the graph with all 450 mappers is completely illegible. Nonetheless, the maximum average read lag for all mappers is about 400 milliseconds.

To test how well the system recuperates from failures, we have enacted a few manual failure scenarios on the streaming processor discussed above.

\begin{figure}[H]
\includegraphics[width=\textwidth]{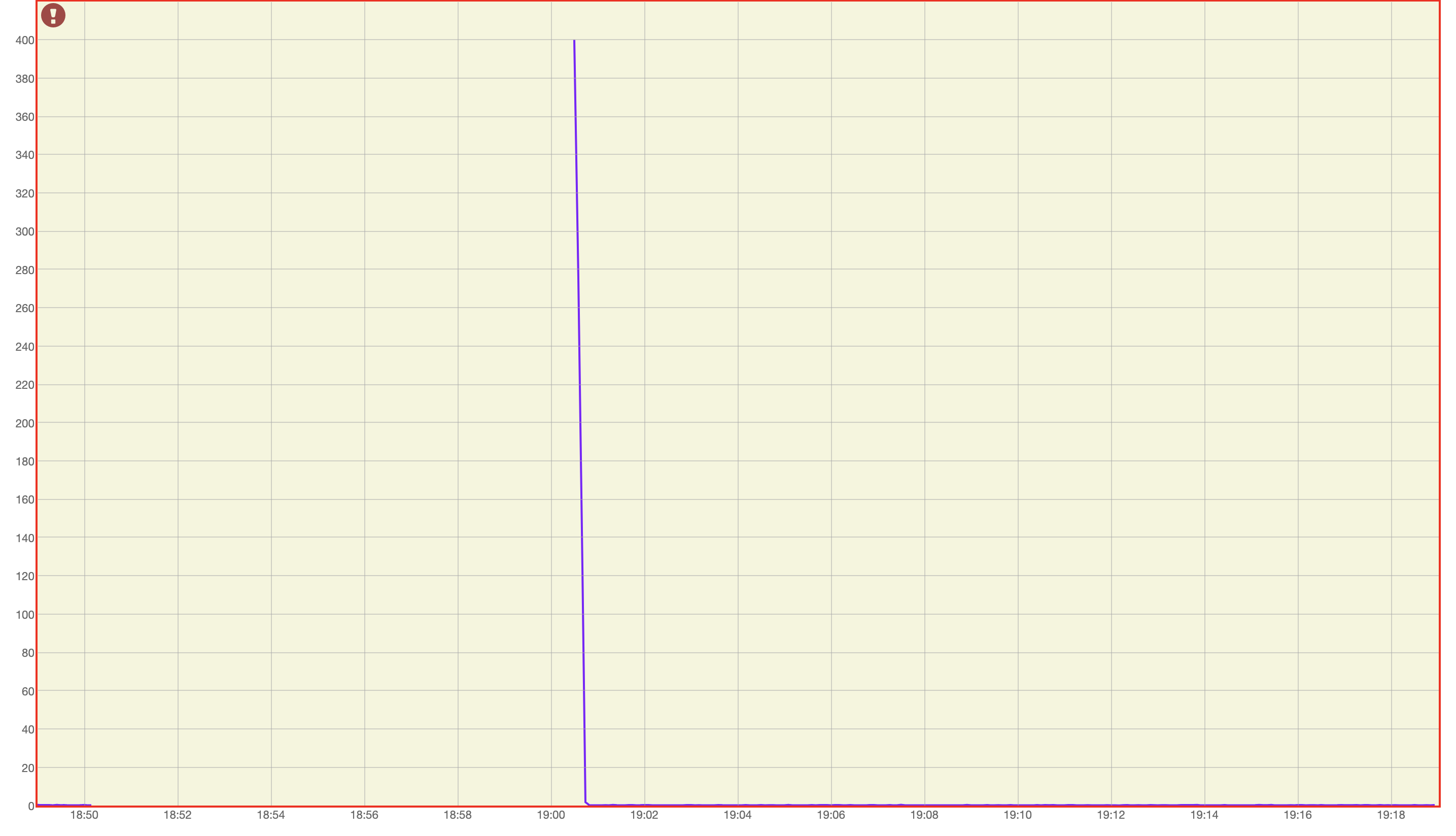}
\centering
\vspace{-1cm}
\caption{A mapper's read lag after a 10 minute failure.}
\label{fig:mapper_failure_read_lag}
\end{figure}

\begin{figure}[H]
\includegraphics[width=\textwidth]{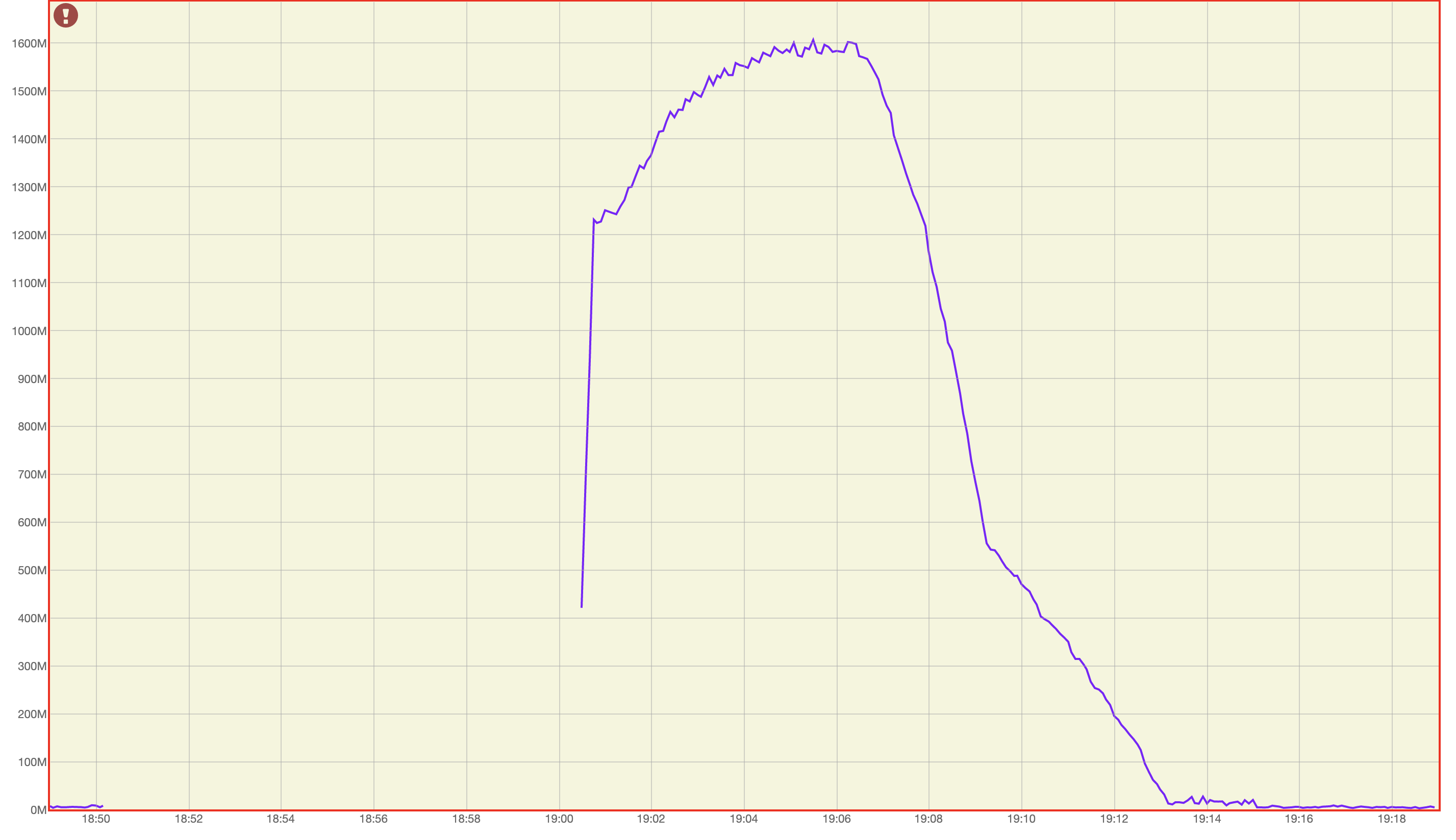}
\centering
\vspace{-1cm}
\caption{A mapper's buffered window size after a 10 minute failure.}
\label{fig:mapper_failure_window_size}
\end{figure}

First, we paused a single mapper for around 10 minutes and killed it at the end of this period, allowing the controller to restart the job. As expected, this didn't cause any reducers to slow down. The important metric here was how fast this mapper could catch up with the stream. As it can be seen from figure \ref{fig:mapper_failure_read_lag} on the next page, the read lag dropped to the same level as it was before the failure in around 15 seconds. This was made possible by the mapper's internal buffer, which temporarily grew to around 1.5 gigabytes out of its 8 gigabyte memory limit, as illustrated in figure \ref{fig:mapper_failure_window_size}. It took around 15 minutes for the mapper to shrink its buffer back to its pre-failure state. In a production scenario one would use more reducers so that their peak processing rate is more substantially higher than the input topic's throughput.


\begin{figure}[H]
\includegraphics[width=\textwidth]{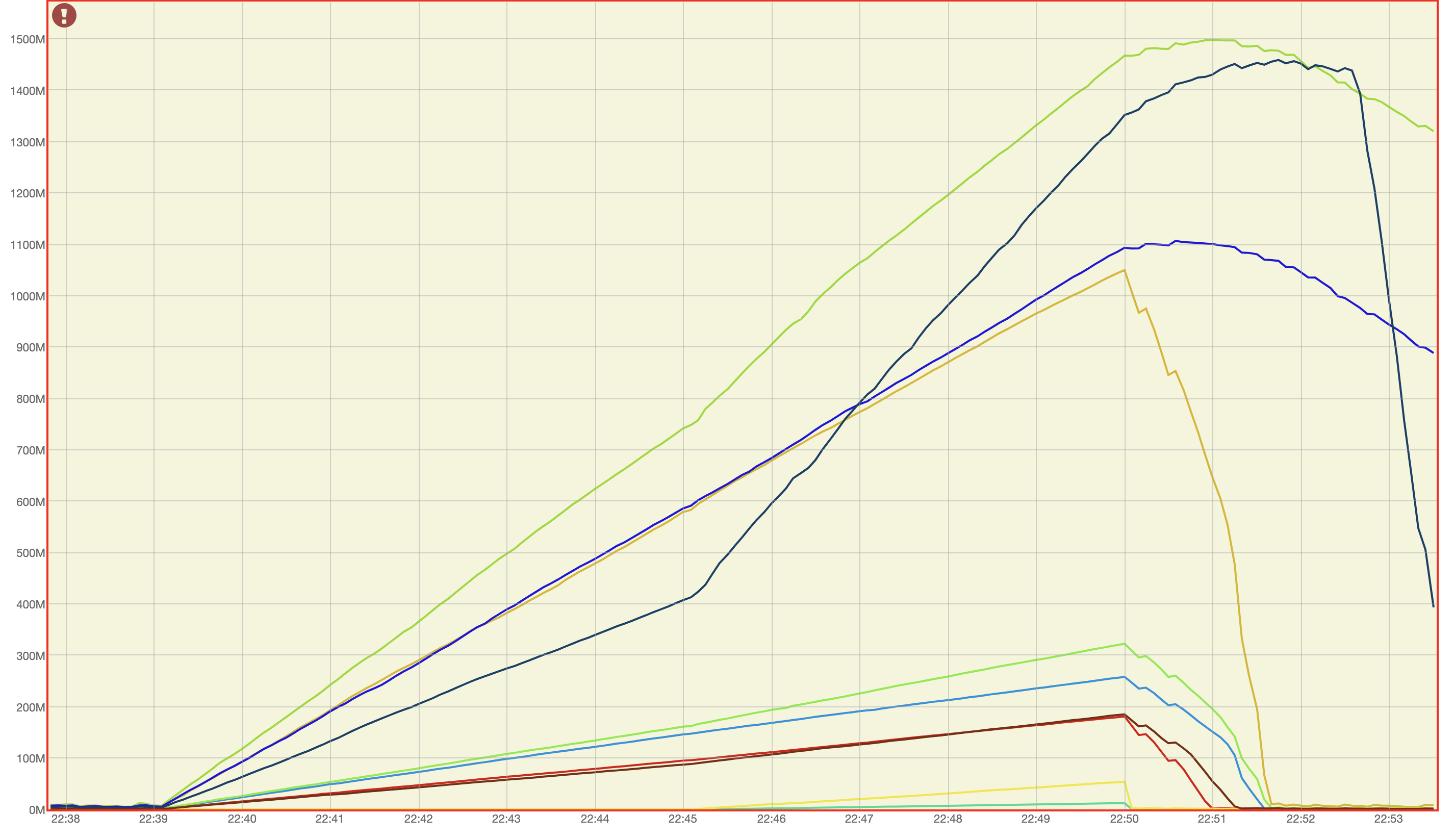}
\centering
\vspace{-1cm}
\caption{Selected mappers' buffered window size during a 10 minute reducer failure.}
\label{fig:reducer_failure_window_size}
\end{figure}

The second scenario we tested was a 10 minute pause and later failure of a single reducer. As discussed previously, our system is not yet well-equipped to handle these kind of downtimes efficiently. Currently, a single unavailable reducer almost completely prevents mappers from trimming their internal buffers. A possible remedy for this drawback is outlined in section \ref{section:fw}. Nonetheless, it is important to assess the current effect reducer halts have on a streaming processor. Predictably, the mappers' buffers grew while the reducer was out, as can be seen in figure \ref{fig:reducer_failure_window_size}. Again, we only present results from 10 evenly selected mappers so that the graph is legible. Once the reducer was back online the mappers quickly recuperated and their windows began shrinking back to their previous sizes in a matter of minutes. Thanks to the mappers' buffers, no other performance metrics were impacted during this test.

From these scenarios it is clear that our system does, in fact, sustain real worker failures and downtimes while maintaining solid processing throughput. The next logical step would have been to test it within a real production workflow, which, sadly, didn't end up fitting in our timeline.

\section{Future work}
\label{section:fw}
There are many directions in which this work can be continued, and some features that will likely be implemented soon. In this chapter we also want to outline the potential designs for some of the proposed functionality, as the ideas behind them are sophisticated enough to be worth mentioning.

Most importantly, to deal with straggling workers, mappers will flush batches and advance their windows when most, but not necessarily all, reducers have processed the rows in these batches. When that happens, rows that are still needed by some reducers will be spilled to a designated table. By configuring thresholds in this approach we will be able to leverage low write amplification factors with sufficient straggler tolerance.

Another goal is to allow a single mapper to read multiple input partitions. It would enable the system to use fewer resources when an input topic has many low-throughput partitions, which causes mappers to be underutilized. The challenge lies in the fact that the order in which data is delivered from distinct partitions is not deterministic. Two batches of rows might be read from two different partitions in one sequence, partially sent off and processed by reducers and then reread in the reverse order if the mapper fails. This would inevitably cause some rows to be lost and others to be processed more than once. To overcome this issue, mappers will read data in one of two modes. In the \textit{advancing} mode a mapper will collect data from its multiple assigned partitions and persist the order and size of the received batches to a tablet of an ordered dynamic table. In the \textit{catch up} mode a mapper will read rows from this tablet and wait to receive the same amount of rows from the corresponding partitions, returning them in exactly the same order. The latter mode will be used when a mapper finds that its state is behind the offsets stored in its designated tablet. Altogether, this would allow us to guarantee that data from multiple partitions will be received by a mapper in a deterministic order, solving the issue in question.

A further limitation of the current system is the reducer interface, which permits working transactionally with only one batch of rows at a time. In this model one cannot perform windowed aggregation while maintaining true exactly-once guarantees. We would like to move to a \textit{persistent queue} interface in which users can request batches of rows as needed, carry out their desired computations and invoke a commit method on batches that have been successfully processed. This method would, naturally, allow a transaction to be passed along, which would be used to update the reducer's persistent state and commit the user's processing side-effects atomically.

To improve processing efficiency it is also possible to modify both the mapper and reducer workflows to run in a pipelined fashion. For example, a single cycle of the reducer's main procedure can be subdivided into three consecutive stages: \textit{fetch}, \textit{process} (combine row batches and run \verb|Reduce|) and \textit{commit}. Thus, we can perform stages within different cycles concurrently, as long as executions of each individual stage are well-ordered. This is a generalization of instruction pipelining utilized in modern processors.

Continuing with client-side features, not all tasks demand strict exactly-once guarantees. For example, jobs calculating generic statistics on a stream of data can usually handle minor losses of rows by reducers and small amounts of data duplication. Thus, we could also want to provide the ability to lift some of the requirements in favor of better processing times.

There is also a long way to go on the usability front. Improvements could include providing an easier way of dealing with \verb|UnversionedRow|s and implementing some common functionality, such as hash partitioning, within designated base classes.

In the long run, our design has the theoretical ability to support snapshotting the state of the system and restarting from said snapshots. It could also be integrated more deeply into a solution which would allow users to run streaming pipelines consisting of several streaming processors. 
\section{Conclusion}
\label{section:conclusion}
Streaming data processing problems arise regularly at IT companies of all scales. It is a vast field of work, with many different solutions available in local and cloud environments. However, there is a lack of suitable systems at Yandex internally. Open-source solutions are usually locked to corresponding underlying infrastructures and cannot be adapted without tremendous overheads. As a result, we have used Yandex's native distributed infrastructure to build an efficient distributed fault-tolerant streaming processing system that achieves low write amplification factors and provides end-to-end exactly-once guarantees.
% Add more results when they are ready.
\printbibliography[heading=bibintoc]
\end{document}

%% file: title.tex
\begin{titlepage}
\newpage

{\setstretch{1.0}
\begin{center}
Federal State Autonomous Educational Institution for Higher Education
National Research University Higher School of Economics
\\
\bigskip
Faculty of Computer Science \\
Applied Mathematics and Information Science \\
\end{center}
}

\vspace{8em}

\begin{center}
{\Large BACHELOR'S THESIS}\\
\textsc{\textbf{
Program project
\linebreak
``Better Write Amplification for Streaming Data Processing''}}
\end{center}

\vspace{4em}

{\setstretch{1.0}
\hfill\parbox{16cm}{
\hspace*{5cm}\hspace*{-5cm}%
Submitted by Chulkov Andrey Sergeevich\\
student of group 175, 4th year of study\\

\hspace*{5cm}\hspace*{-5cm}%
Approved by Supervisor:\\
Akhmedov Maxim Basirovich\\

}
}

\vspace{\fill}

\begin{center}
Moscow 2021
\end{center}

\end{titlepage}